\documentclass[a4paper,aps,prd,10pt,preprintnumbers,showpacs,twocolumn,superscriptaddress,nofootinbib,amsmath,amssymb]{revtex4-1}
\usepackage{graphicx}
\usepackage[utf8]{inputenc}
\usepackage[T1]{fontenc}
\usepackage{cmap}

\def\imo{i}

\def\K{{\cal K}}

\newcommand{\ie}{{i.e.,}~}
\newcommand{\eg}{{e.g.,}~}

\begin{document}

\title{Quasinormal ringing of general spherically symmetric parametrized black holes}
\author{R. A. Konoplya} \email{roman.konoplya@gmail.com}
\affiliation{Research Centre for Theoretical Physics and Astrophysics, Institute of Physics, Silesian University in Opava, Bezručovo náměstí 13, CZ-74601 Opava, Czech Republic}

\author{A. Zhidenko}\email{olexandr.zhydenko@ufabc.edu.br}
\affiliation{Centro de Matemática, Computação e Cognição (CMCC), Universidade Federal do ABC (UFABC), \\ Rua Abolição, CEP: 09210-180, Santo André, SP, Brazil}

\begin{abstract}
The general parametrization of spherically symmetric and asymptotically flat black-hole spacetimes in arbitrary metric theories of gravity was suggested in \cite{Rezzolla:2014mua}. The parametrization is based on the continued fraction expansion in terms of the compact radial coordinate and has superior convergence and strict hierarchy of parameters. It is known that some observable quantities, related to particle motion around the black hole, such as the eikonal quasinormal modes, radius of the shadow, frequency at the innermost stable circular orbit, and others, depend mostly on only a few of the lowest coefficients of the parametrization. Here we continue this approach by studying the dominant (low-lying) quasinormal modes for such generally parametrized black holes. We show that, due to the hierarchy of parameters, the dominant quasinormal frequencies are also well determined by only the first few coefficients of the expansion for the so-called \emph{moderate} black-hole geometries. The latter are characterized by a relatively slow change of the metric functions in the radiation zone near the black hole. The nonmoderate metrics, which change strongly between the event horizon and the innermost stable circular orbit are usually characterized by echoes or by the distinctive (from the Einstein case) quasinormal ringing which does not match the current observational data. Therefore, the compact description of a black-hole spacetime in terms of the truncated general parametrization is an effective formalism for testing strong gravity and imposing constraints on allowed black-hole geometries.
\end{abstract}
\maketitle

\section{Introduction}

The near-future observations of black holes in the gravitational and electromagnetic spectra should allow us to test Einstein theory and its alternatives in the strong-gravity regime via determining the black-hole geometry \cite{alternative1}. At the same time, the current observational data still leave a great deal of room for deviations from Einstein gravity \cite{alternative2}. In order to avoid the consideration of various astrophysical phenomena in each theory of gravity, case by case, and have a universal and powerful formalism for comparison of the experimental data with theoretical predictions, the general parametrization of spherically symmetric and asymptotically flat black-hole spacetime was developed in \cite{Rezzolla:2014mua} and extended to the axial symmetry in \cite{Konoplya:2016jvv}. This parametrization is based on the double expansion in the form of the infinite continued fraction in terms of the compact radial coordinate and respectively the equatorial plane. The expansion is similar in spirit to the parametrized post-Newtonian (PPN) formalism, but valid in the whole space outside the event horizon up to spatial infinity. It possesses superior convergence and a strict hierarchy of parameters, so that constraining the parameters by observations should tell which theory of gravity is closer to the given experimental data.

The parametrization formalism of \cite{Rezzolla:2014mua,Konoplya:2016jvv} has been broadly used for finding various analytical black-hole metrics \cite{Kokkotas:2017zwt} which serve as analytic approximations to the solutions obtained numerically. Using this approach to describe the black-hole geometry a number of phenomena in the background of these parametrized black-hole metrics, such as quasinormal modes (QNMs) \cite{reviews}, particle motion, Hawking radiation \cite{Hawking:1975vcx}, shadows, and the Blandford-Znajek effect have been studied in \cite{pappl}. The initial parametrization was generalized to the case of higher-dimensional black holes \cite{Konoplya:2020kqb} and to four-dimensional wormhole spacetimes \cite{Bronnikov:2021liv}.

The general parametrization \cite{Rezzolla:2014mua,Konoplya:2016jvv} contains an infinite number of parameters. It was shown in \cite{Konoplya:2020hyk} that some astrophysically relevant quantities such as the eikonal quasinormal modes, the radius of the shadow, the frequency at the innermost stable circular orbit etc., must depend mostly on a few of these parameters -- at least for a broad class of black-hole geometries called \emph{moderate}. A moderate black-hole metric implies that the metric functions do not change quickly within the radiation zone, \ie between the event horizon and the innermost stable circular orbit. In other words the geometry goes over into the asymptotic regime (described via post-Newtonian approximations) relatively slowly, as it occurs for the Schwarzschild black hole and its analogs in various alternative theories of gravity. This concept is compatible with our understanding that, in a true theory of gravity, the observable quantities should not deviate from their Schwarzschild values by orders, but rather, at most by tens of percent. Otherwise, such strong deviations would be observable within the post-Newtonian regime.

The opposite, \emph{nonmoderate} black-hole metric can be indistinguishable from the Kerr form in the whole space, except for a very small region near its horizon, where the deviation is huge. Then, such a geometry would also be experimentally indistinguishable from the Kerr one, leaving a weak imprint only in the form of gravitational echoes at late times, when the signal is almost completely damped \cite{Cardoso:2016rao}.

When considering a parametrized approximate metric instead of some exact black-hole solution (once the latter is numerical), the criterium of sufficient accuracy of the approximation is evident: the physical ``effect'', which is the deviation of some measurable physical quantities from their Schwarzschild values, must be at least one order larger than the relative error of the approximation due to the truncation of the parametrization. Using the eikonal characteristics of spherically symmetric black holes (such as eikonal quasinormal modes, the radius of the shadow and the frequency at the innermost stable circular orbit) we have shown in \cite{Konoplya:2020hyk} that moderate metrics can be very well described by only three parameters in this case, and owing to the strong hierarchy of coefficients, five parameters are sufficient if one order higher accuracy is necessary.

It is known that the quasinormal modes in the regime of high multipole numbers (eikonal regime) are linked to the parameters of the null geodesics \cite{Cardoso:2008bp} and this link, although not obligatory for gravitational and other nonminimally coupled field, is guaranteed for test fields \cite{Konoplya:2017wot}, once the black hole is spherically symmetric and asymptotically flat or de Sitter. Therefore, a number of phenomena, such as the radius of shadows, characteristics of particle motion and accretion are intrinsically linked to the eikonal quasinormal modes as well. In this context, despite a number of papers on testing of the parametrized black holes \cite{Carson:2020iik}, no convincing work has been done so far for the characteristics beyond the eikonal regime, that are not connected to particle motion. First of all, no such analysis was suggested for the low-lying quasinormal modes, which are especially important because they dominate in a signal. The analysis of quasinormal modes in \cite{Volkel:2019muj} was aimed at the attempt to solve the inverse problem: from the quasinormal spectrum to parametrization, which evidently could not be effectively solved via determining only the value of the dominant mode.

In the present paper we will consider the dominant quasinormal modes for a general parametrized black hole of \cite{Rezzolla:2014mua} and show that the strict hierarchy of parameters is indeed present: the low-lying quasinormal modes strongly depend on the lower coefficients and much less on the higher ones. Therefore, the dominant quasinormal modes of an asymptotically flat spherically symmetric black hole essentially depend upon only three first coefficients of the parametrization. Higher coefficients can only slightly correct quasinormal modes. The exception from this picture is provided by nonmoderate metrics which either do not satisfy the constrains of the post-Newtonian regime or have so strong a deviation from the Schwarzschild quasinormal ringing that it would immediately be seen in the current experiments.

Here we use the approach of parametrization of the metric and not of the effective potential used in \cite{McManus:2019ulj}. Unfortunately, this approach does not allow us to find gravitational quasinormal modes for the general case unless the underlying gravitational theory is specified. It was shown in \cite{Suvorov:2021amy} that it is possible to find a mixed scalar-$f(R)$ theory in which the given metric is an exact solution. However, for the general static spherically symmetric black hole in the corresponding theory, the equations for axial gravitational perturbations, which do not couple to the scalar-field degree of freedom, still depend on a free parameter of the theory. In order to avoid the ambiguity, in the present paper we study quasinormal modes of test fields, which usually are qualitatively similar to the gravitational ones and approach the latter very quickly when the multipole number is increased.

The paper is organized as follows. In Section~\ref{sec:param} we briefly review the general parametrization of \cite{Rezzolla:2014mua} and suggest some basic constraints on the values of the parametrization coefficients. Section~\ref{sec:eqs} introduces the master wave equations and the methods used for calculations of quasinormal modes. Section~\ref{sec:moderate} is devoted to the quasinormal modes of moderate black holes, while Section~\ref{sec:nonmoderate} considers quasinormal ringing and echoes which take place for nonmoderate black-hole geometries. Finally, in the Conclusions we summarize the obtained results and mention some open questions.

\section{The parametrized black-hole metric}\label{sec:param}

The metric of a spherically symmetric black hole can be written in the following general form,
\begin{equation}
ds^2=-N^2(r)dt^2+\frac{B^2(r)}{N^2(r)}dr^2+r^2 (d\theta^2+\sin^2\theta d\phi^2),\label{metric}
\end{equation}
where $r_0$ is the event horizon, so that
\begin{equation}
N(r_0)=0.
\end{equation}
Following \cite{Rezzolla:2014mua}, we will use the new dimensionless variable
\begin{equation}
x \equiv 1-\frac{r_0}{r},
\end{equation}
so that $x=0$ corresponds to the event horizon, while $x=1$ corresponds to spatial infinity. In addition, we rewrite the metric function $N$ as follows:
\begin{equation}
N^2=x A(x),
\end{equation}
where $A(x)>0$ for \mbox{$0\leq x\leq1$}. Using the new parameters $\epsilon$, $a_0$, and $b_0$, the functions $A$ and $B$ can be written as
\begin{eqnarray}\nonumber
A(x)&=&1-\epsilon (1-x)+(a_0-\epsilon)(1-x)^2+{\tilde A}(x)(1-x)^3\,,
\\
B(x)&=&1+b_0(1-x)+{\tilde B}(x)(1-x)^2\,.\label{ABexp}
\end{eqnarray}
Here the coefficient $\epsilon$ measures the deviation of $r_0$ from the Schwarzschild radius $2 M$,
$$\epsilon = \frac{2 M-r_0}{r_0}.$$

The coefficients $a_0$ and $b_0$ can be considered as combinations of the PPN parameters,
$$a_0=\frac{(\beta-\gamma)(1+\epsilon)^2}{2}, \qquad b_0=\frac{(\gamma-1)(1+\epsilon)}{2}.$$
Current observational constraints on the PPN parameters imply \mbox{$a_0 \sim b_0 \sim 10^{-4}$}, so that from here and on we will consider them as null.

The functions ${\tilde A}$ and ${\tilde B}$ are introduced through infinite continued fraction in order to describe the metric near the horizon (\ie for $x \simeq 0$),
\begin{equation}\label{ABdef}
{\tilde A}(x)=\frac{a_1}{\displaystyle 1+\frac{\displaystyle a_2x}{\displaystyle 1+\ldots}}, \qquad
{\tilde B}(x)=\frac{b_1}{\displaystyle 1+\frac{\displaystyle b_2x}{\displaystyle1+\ldots}},
\end{equation}
where $a_1, a_2,\ldots$ and $b_1, b_2,\ldots$ are dimensionless constants to be constrained from observations of phenomena which are localized near the event horizon. At the horizon only the first term in each of the continued fractions survives,
$
{\tilde A}(0)={a_1},~
{\tilde B}(0)={b_1},
$
which implies that near the horizon only the lower-order terms of the expansions are essential.

The parametrization coefficient $\epsilon$ is fixed as follows:
\begin{equation}
\epsilon=\frac{2M}{r_0}-1\equiv2C-1,
\end{equation}
where the ratio $C\equiv M/r_0$ is called the \emph{compactness}.

Since the Kerr-Newman black-hole compactness obeys the inequality
\begin{equation}
1/2\leq C < 1,
\end{equation}
the range of values for $\epsilon$ in General Relativity is
\begin{equation}
0 \leq \epsilon < 1.
\end{equation}

At the same time, the experimental data suggests existence of the neutron stars with the compactness $C\approx0.2$ ($M\approx1.5M_{\odot}$, $R\approx11~\textrm{km}$) \cite{Ozel:2015fia}. This imposes a lower bound on the allowed values of $\epsilon$: at least, for the highly-rotating black holes,
\begin{equation}
\epsilon\gtrsim-0.6.
\end{equation}
Otherwise, if a black hole with such compactness could exist, the corresponding neutron stars would collapse. One could assume the existence of some phenomenon that prevents the neutron star from collapsing even though its compactness is higher than the black-hole one. Yet, it is unnatural to expect that such a hypothetical mechanism changes the above bound significantly.

Since the angular momentum of a neutron star can prevent its collapse, the above bounds are, strictly speaking, valid for the highly-rotating black holes only. However, it is natural to assume that the compactness of slowly-rotating and nonrotating black holes are of the same order of magnitude. The additional assumption that the star collapsing into a black hole further increases the compactness and the lower bound for $\epsilon$. Assuming existence of the stable maximum-mass neutron stars, which can have the compactness $C\approx0.3$ ($M\approx2M_{\odot}$, $R\approx10~\textrm{km}$) \cite{Steiner:2010fz}, increases the above bound to $$\epsilon\gtrsim-0.4.$$

At the same time the upper bound for the values of $\epsilon$, i.~e. maximally allowed compactness of the black hole, seems to be impossible to estimate. In Einstein gravity there is the uniqueness theorem claiming that the Schwarzschild/Kerr solution is the only external geometry for the black hole \cite{Israel:1967wq}. However, it is not guaranteed that such uniqueness will be provided in all alternative theories of gravity. In other words, we cannot exclude a possibility to have two black holes of the same size but with different masses, where different black-hole solutions were realized during the formation of black holes. Such nonuniqueness occurs, for example, in the Einstein-Weyl theory. In this case a more massive black hole is described by a higher value of $\epsilon$. After all, because of enormous distances to black holes, the current experimental data do not allow us to constrain the radius of them even up to some reasonable margins.

Substituting the above expression (\ref{ABdef}) for ${\tilde A}(x)$ into (\ref{ABexp}), we find that the expansion of $A(x)$ has the form:
$$A(x)=1-\epsilon (1-x)+(a_0-\epsilon)(1-x)^2+\frac{(1-x)^3a_1}{\displaystyle 1+\frac{\displaystyle a_2x}{\displaystyle 1+\ldots}}\,.$$
Then, assuming that the surface gravity must be positive at the event horizon, we have the following bound:
\begin{equation}\label{upbound}
\frac{dN^2(0)}{dx}=A(0)=1-2\epsilon+a_0+a_1>0.
\end{equation}

In the following subsections we will show that the low-lying quasinormal modes of moderate black-hole geometries are well determined by only the three coefficients of the parametrization. In this case, the metric functions are
\begin{eqnarray}\nonumber
N^2(r)&=&1 - \frac{r_{0}(\epsilon+1)}{r} + \frac{r_{0}^3(\epsilon+a_1)}{r^3} - \frac{r_0^4 a_{1}}{r^4},\\\label{metric1}
B^2(r)&=&\left(1+\frac{r_0^2 b_{1}}{r^2}\right)^2.
\end{eqnarray}
The parameters $\epsilon$, $a_1$ and $b_1$ are such that when they all are equal to zero, the Schwarzschild limit is reproduced. Within the approximation~(\ref{metric1}) the deviation of observable quantities are at least one order larger than the relative error. For a more accurate approximation, such that the error is two orders smaller than the ``effect'', one can use the expansion~(\ref{ABdef}) to include higher coefficients, $a_2$ and $b_2$, in the metric functions.

\section{The master wave equations and the methods for calculations of the quasinormal modes}\label{sec:eqs}

The general covariant equations for the test scalar $\Phi$ and electromagnetic $A_\mu$ fields in the black-hole background have the following form,
\begin{equation}\label{perturbeqs}
\begin{array}{rcl}
\dfrac{1}{\sqrt{-g}}\partial_\mu \left(\sqrt{-g}g^{\mu \nu}\partial_\nu\Phi\right)&=&0,\\
\dfrac{1}{\sqrt{-g}}\partial_{\mu} \left(F_{\rho\sigma}g^{\rho \nu}g^{\sigma \mu}\sqrt{-g}\right)&=&0,
\end{array}
\end{equation}
where $F_{\mu\nu}=\partial_\mu A_\nu-\partial_\nu A_\mu$.
After the separation of variables Eqs.~(\ref{perturbeqs}) can be reduced to the Schrödinger-like form (see, \eg \cite{reviews}),
\begin{equation}\label{wave-equation}
\dfrac{\partial^2 \Psi}{\partial t^2}-\dfrac{\partial^2 \Psi}{\partial r_*^2}+V(r)\Psi=0,
\end{equation}
where the ``tortoise coordinate'' $r_*$ is defined by the relation
\begin{equation}
dr_*=\frac{B(r)}{N^2(r)}dr.
\end{equation}

The effective potentials for the scalar and electromagnetic fields are
\begin{equation}\label{potentialScalar}
V(r)=N^2(r)\frac{\ell(\ell+1)}{r^2} + \frac{1-s}{2r}\frac{d}{dr}\frac{N^4(r)}{B^2(r)},
\end{equation}
where $\ell=1, 2, \ldots$ are the multipole numbers and $s=0$ ($s=1$) corresponds to the scalar (electromagnetic) field, respectively. The effective potential for the electromagnetic field has the form of the positive definite potential barrier, while this is not always so for a scalar field.

Quasinormal modes $\omega_{n}$ are frequencies corresponding to solutions of the master wave equation (\ref{wave-equation}) with the requirement of the purely outgoing waves at infinity and at the event horizon,
\begin{equation}
\Psi \propto e^{-\imo \omega t \pm \imo \omega r_*}, \quad r_* \to \pm \infty.
\end{equation}

In order to find low-lying quasinormal modes we will use the two methods: the time-domain integration method and the WKB method.

In the time domain, we can integrate the wavelike equation (\ref{wave-equation}) in terms of the light-cone variables $u=t-r_*$ and $v=t+r_*$.
We will use the discretization scheme proposed in \cite{Gundlach:1993tp},
\begin{eqnarray}\label{Discretization}
\Psi\left(N\right)&=&\Psi\left(W\right)+\Psi\left(E\right)-\Psi\left(S\right)\\\nonumber
&&-\frac{\Delta^2}{4}V\left(S\right)\left(\Psi\left(W\right)+\Psi\left(E\right)\right)+{\cal O}\left(\Delta^4\right)\,,
\end{eqnarray}
where we introduced the following notations for the points:
$N\equiv\left(u+\Delta,v+\Delta\right)$, $W\equiv\left(u+\Delta,v\right)$, $E\equiv\left(u,v+\Delta\right)$, and $S\equiv\left(u,v\right)$.
The Gaussian initial data are imposed on the two null surfaces, $u=u_0$ and $v=v_0$. Then, the dominant quasinormal frequencies can be extracted from the time-domain profiles imagined as a sum of exponents with the help of the Prony method \cite{Prony}.

In the frequency domain we will use the WKB method suggested in \cite{Schutz:1985zz} and extended to higher orders in \cite{Iyer:1986np,Konoplya:2003ii,Matyjasek:2017psv}. The method achieved even higher accuracy via the use of the Padé approximants \cite{Matyjasek:2017psv,Hatsuda:2019eoj}.
The higher-order WKB formula has the following form \cite{Konoplya:2019hlu},
\begin{eqnarray}\label{WKBformula}
&&\omega^2=V_0+A_2(\K^2)+A_4(\K^2)+A_6(\K^2)+\ldots\\\nonumber&&-\imo \K\sqrt{-2V_2}\left(1+A_3(\K^2)+A_5(\K^2)+A_7(\K^2)\ldots\right),
\end{eqnarray}
where $\K$ is half-integer. The corrections $A_k(\K^2)$ to the eikonal formula are of the order $k$ and polynomials in $\K^2$ with rational coefficients, which depend on the values of higher derivatives of the potential $V(r)$ at its maximum. In order to increase the accuracy of the WKB formula, we will follow the Matyjasek-Opala approach \cite{Matyjasek:2017psv} and use the Padé approximants. Here we will use the sixth-order WKB method with $\tilde{m}=5$ (where $\tilde{m}$ is defined in \cite{Matyjasek:2017psv,Konoplya:2019hlu}), because this choice provides the best accuracy in the Schwarzschild limit (when all the expansion coefficients are zero) and a good concordance with the results obtained from the time-domain integration.

\section{Quasinormal ringing of moderate black holes}\label{sec:moderate}

From the constraints discussed in section~\ref{sec:param} we see that the values of the parameter $\epsilon$ must not be smaller than $-0.6$.
It follows from (\ref{upbound}) that, for the spherically symmetric black holes, it cannot be larger than $0.5$ if $a_0=a_1=0$, because at larger values no event horizon exists. If the black-hole geometry is moderate, \ie the metric functions change sufficiently slowly in the radiation zone, the higher coefficients of the parametrization cannot be much larger then the lower ones. In that case the upper bound on $\epsilon$ does not change much. Therefore, we will constrain it here by the range
\begin{equation}
-0.5 \leq \epsilon \leq 0.5.
\end{equation}

\begin{figure*}
\resizebox{\linewidth}{!}{\includegraphics*{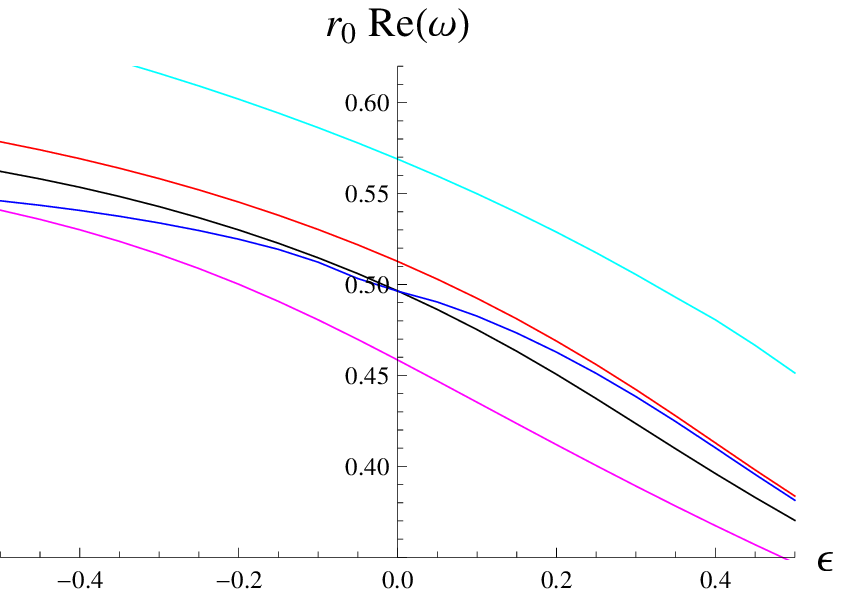}\includegraphics*{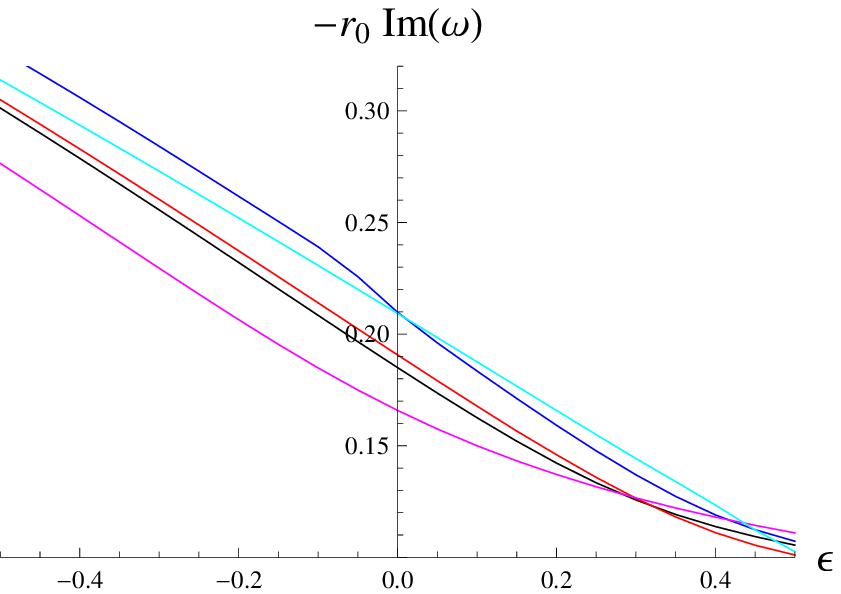}}
\resizebox{\linewidth}{!}{\includegraphics*{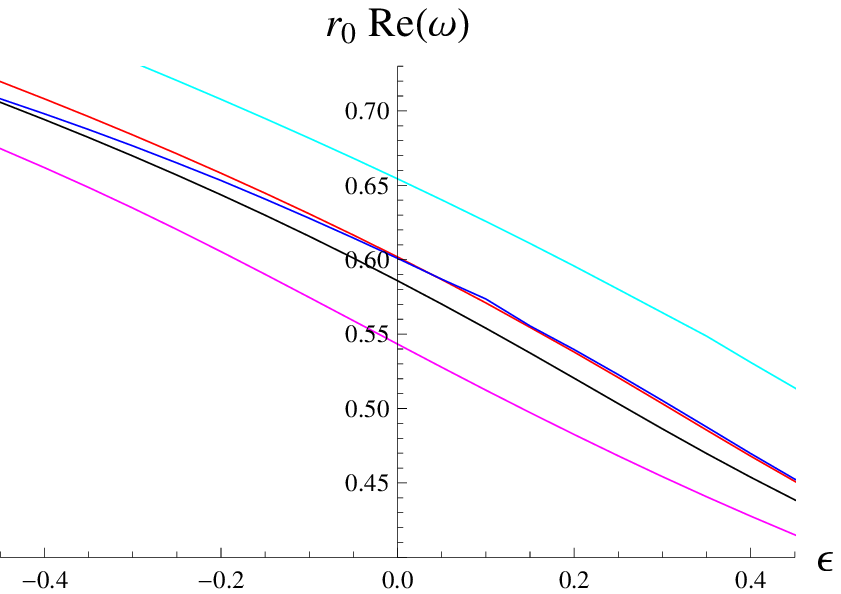}\includegraphics*{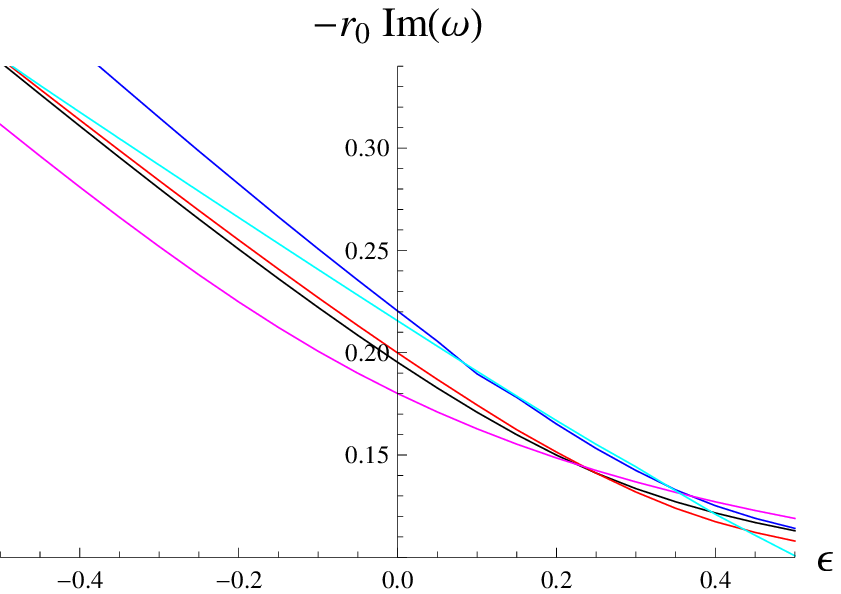}}
\caption{Real and imaginary parts of the fundamental quasinormal mode ($\ell=1$) of the electromagnetic ($s=1$, top panels) and scalar field ($s=0$, bottom panels) as a function of $\epsilon$ for various $a_{1}$ and $b_{1}$: $a_{1} =0.9$, $b_{1}=0.5$ (cyan); $a_{1}=0.1$, $b_{1}=0.2$ (red); $a_{1} =0.2$, $b_{1}=-0.1$ (blue); $a_{1}=b_{1}=0$ (black); $a_{1} =-0.5$, $b_{1}=-0.1$ (magneta).}\label{fig:1}
\end{figure*}

From Fig.~\ref{fig:1} we see that the parameter of deviation of the black-hole radius from its Schwarzschild value is the most important parameter: quasinormal frequencies depend strongly on it and may vary by quite a few tens of percent. The larger $\epsilon$ is, the smaller real oscillation frequency and damping rate is. Thus, black holes, which are more compact than the Schwarzschild one, have smaller oscillation frequencies which damp for a longer time. Since for the moderate black holes the dependence of the quasinormal modes on the parametrization coefficients is similar for $s=0$ and $s=1$, we further consider the quasinormal modes of the electromagnetic field ($s=1$).

\begin{figure*}
\resizebox{\linewidth}{!}{\includegraphics*{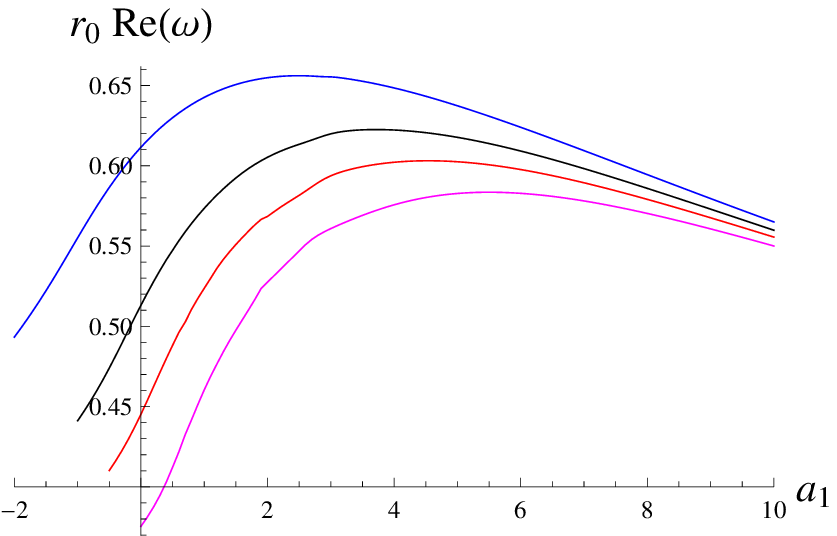}\includegraphics*{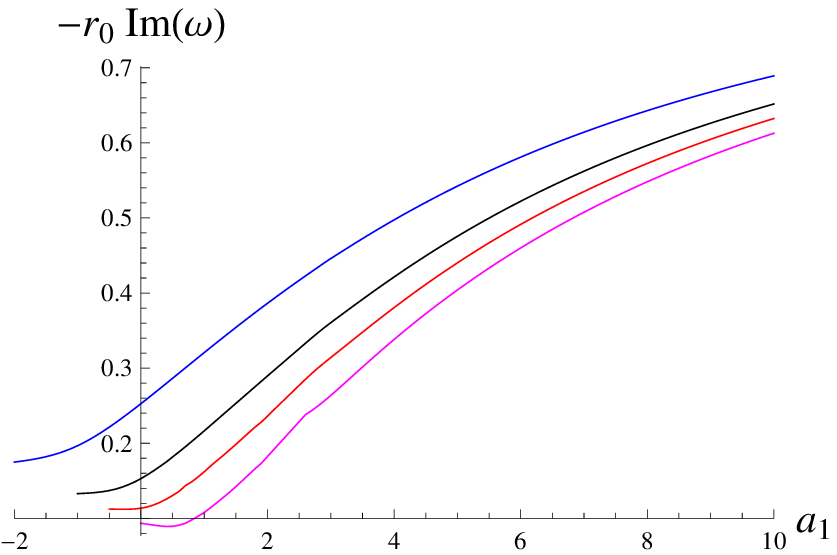}\includegraphics*{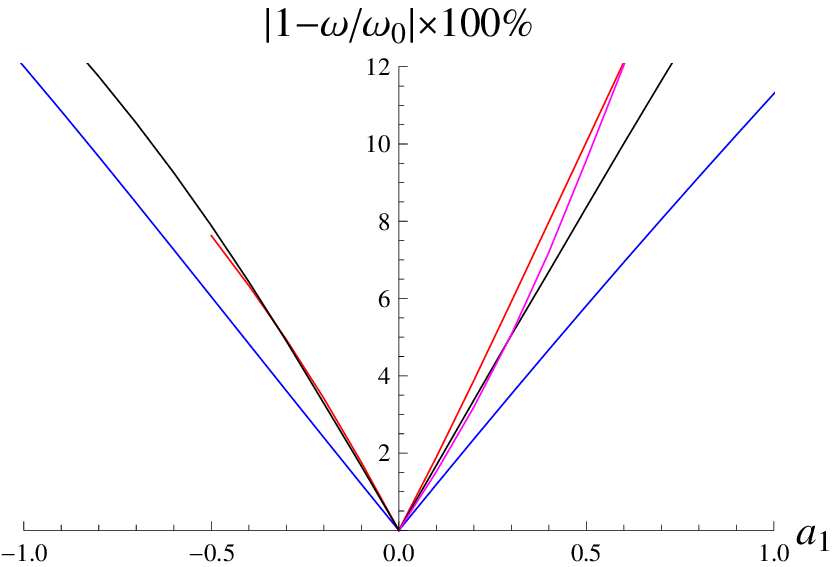}}
\caption{Real and imaginary parts of the fundamental quasinormal mode ($s=1$, $\ell=1$, $n=0$) for $b_{1}=0.5$ and various $\epsilon$ as a function of $a_1$, and the relative deviation from $\omega_0$, corresponding to $a_1=0$: $\epsilon=-0.5$ (blue); $\epsilon=0$ (black); $\epsilon=0.25$ (red); $\epsilon=0.5$ (magenta).}\label{fig:2}
\end{figure*}

\begin{figure*}
\resizebox{\linewidth}{!}{\includegraphics*{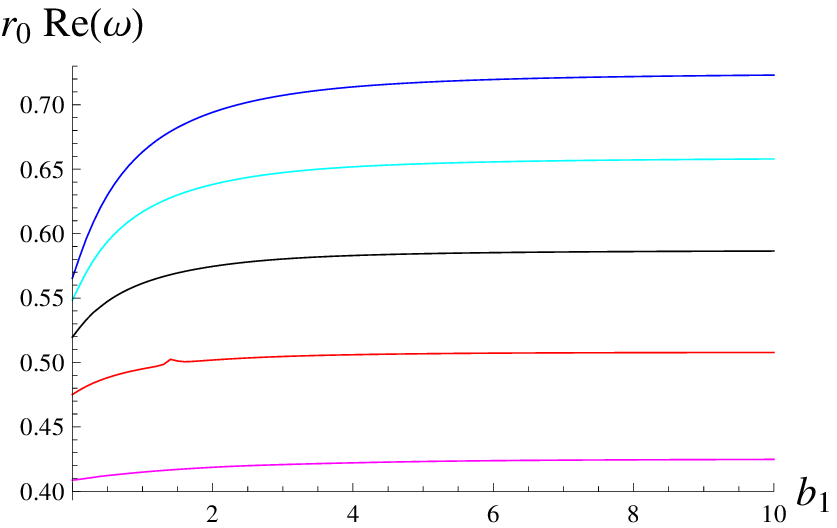}\includegraphics*{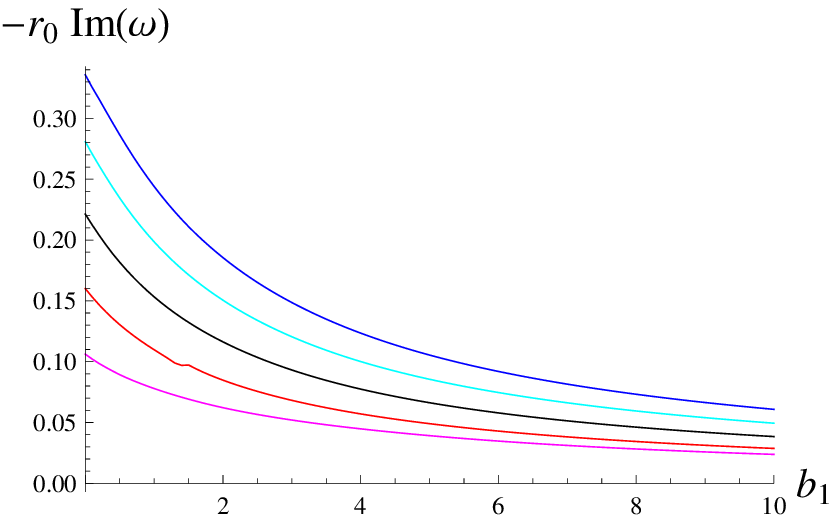}\includegraphics*{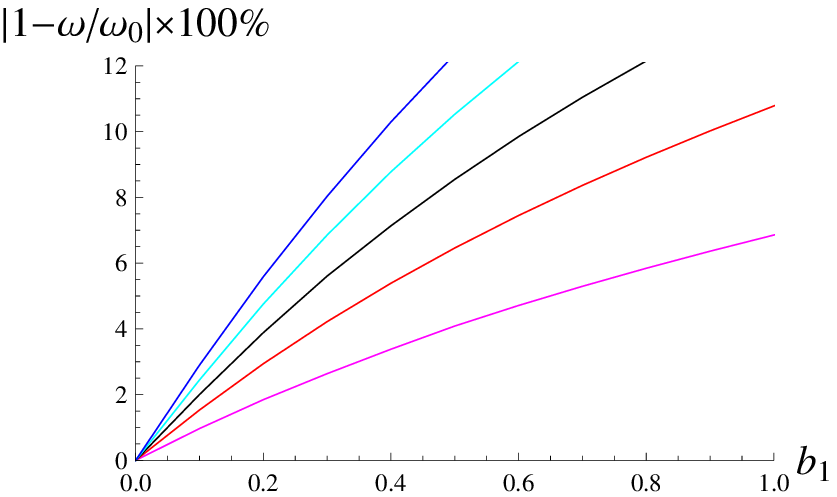}}
\caption{Real and imaginary parts of the fundamental quasinormal mode ($s=1$, $\ell=1$, $n=0$) for $a_{1}=0.5$ and various $\epsilon$ as a function of $b_1$, and the relative deviation from $\omega_0$, corresponding to $b_1=0$: $\epsilon=0.5$ (magenta); $\epsilon=0.25$ (red); $\epsilon=0$ (black); $\epsilon=-0.25$ (cyan); $\epsilon=-0.5$ (blue).}\label{fig:3}
\end{figure*}

From Figs.~\ref{fig:2}~and~\ref{fig:3} we can see that once the higher coefficients $a_1$ and $b_1$ do not exceed $\epsilon$ by an order which is necessary to keep the metric moderate -- they change the quasinormal frequencies only by several percent. Even if we do not limit these coefficients we see that the quasinormal modes quickly approach the asymptotic regime when further increasing of $a_1$ and $b_1$ does not change the quasinormal modes. On the other hand, large negative values of $a_1$ are excluded by the inequality (\ref{upbound}), and negative values of $b_1$ usually correspond to the nonmoderate geometry, which we discuss in the next section.

\begin{figure*}
\resizebox{\linewidth}{!}{\includegraphics*{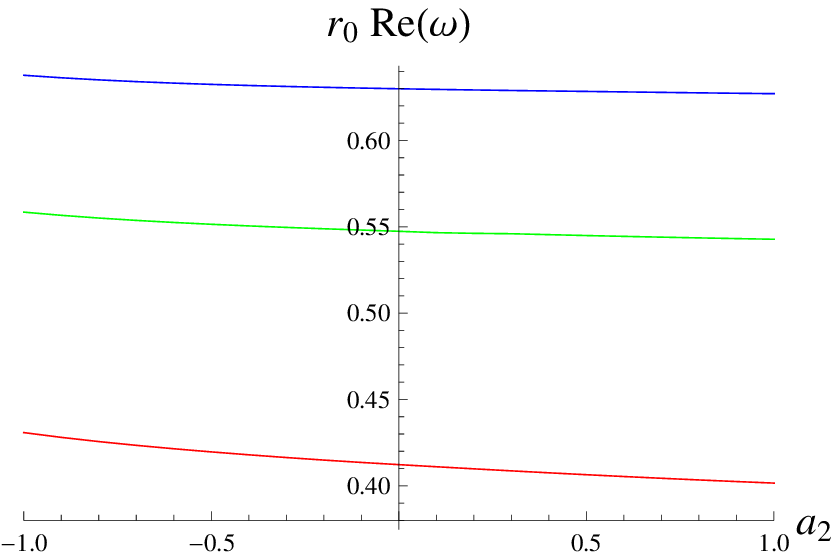}\includegraphics*{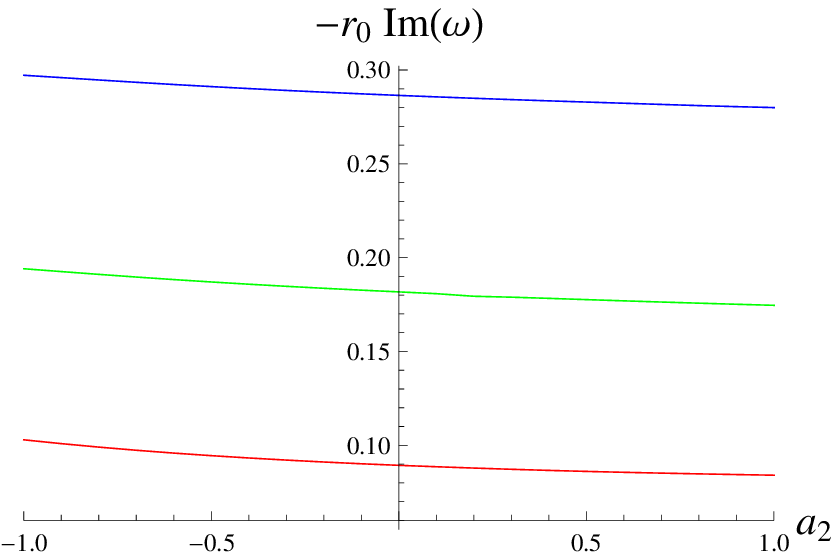}\includegraphics*{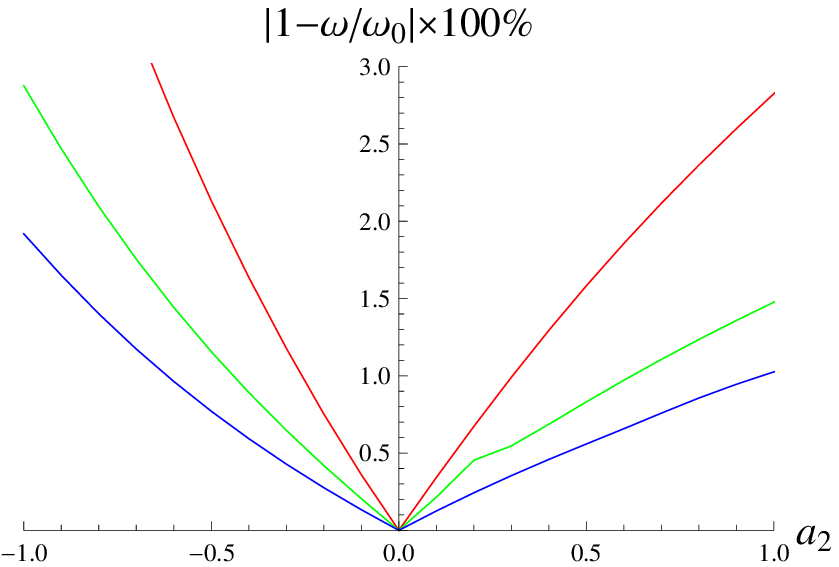}}
\resizebox{\linewidth}{!}{\includegraphics*{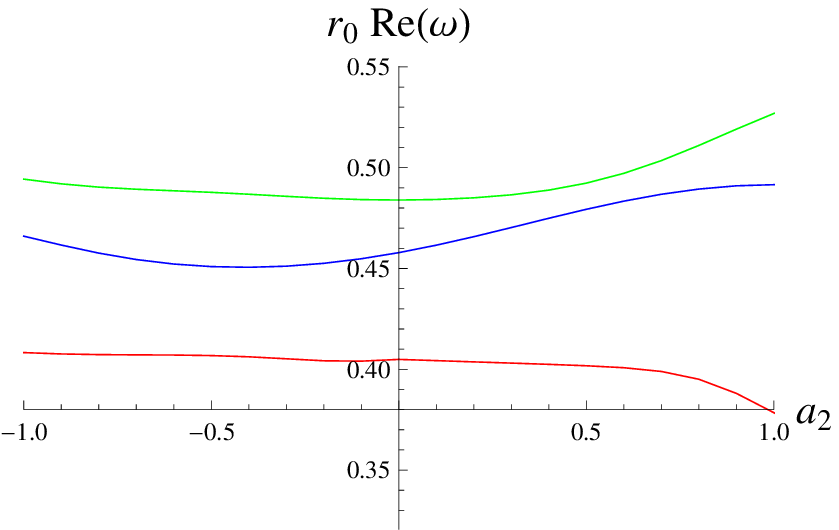}\includegraphics*{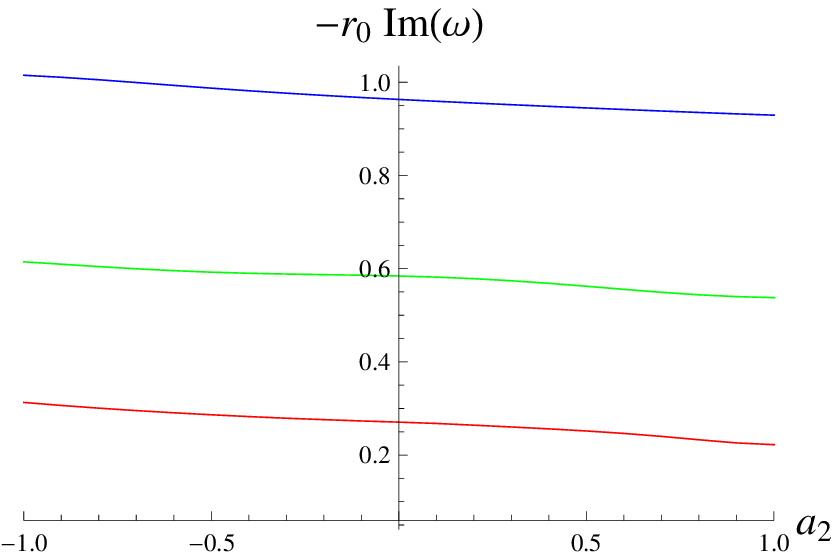}\includegraphics*{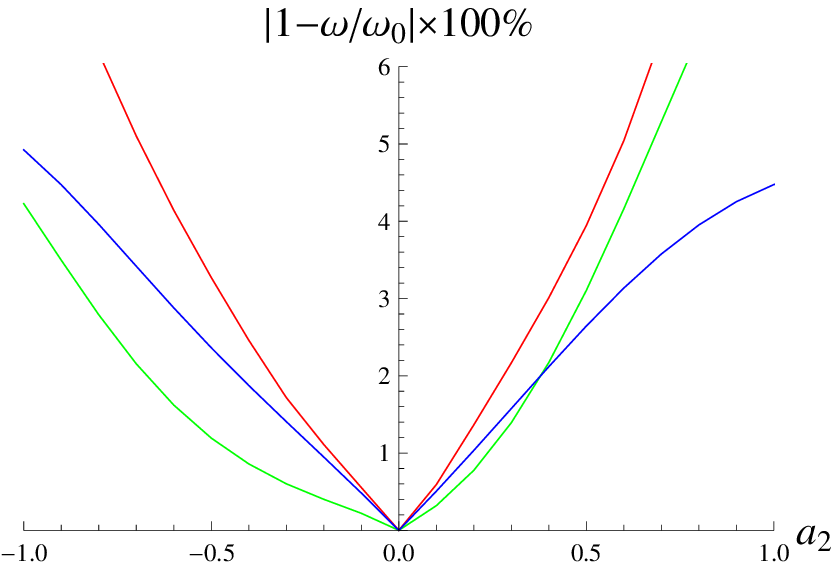}}
\caption{Real and imaginary parts of the fundamental quasinormal mode ($n=0$, top panels) and the first overtone ($n=1$, bottom panels) of the electromagnetic field ($s=1$, $\ell=1$) for $a_{1}=b_{1}=0.5$, $b_2 =0$ and various $\epsilon$ as a function of $a_2$, and the relative deviation from $\omega_0$, corresponding to $a_2=0$: $\epsilon=0.5$ (red), $\epsilon=0$ (green), $\epsilon=-0.5$ (blue).}\label{fig:4}
\end{figure*}

\begin{figure*}
\resizebox{\linewidth}{!}{\includegraphics*{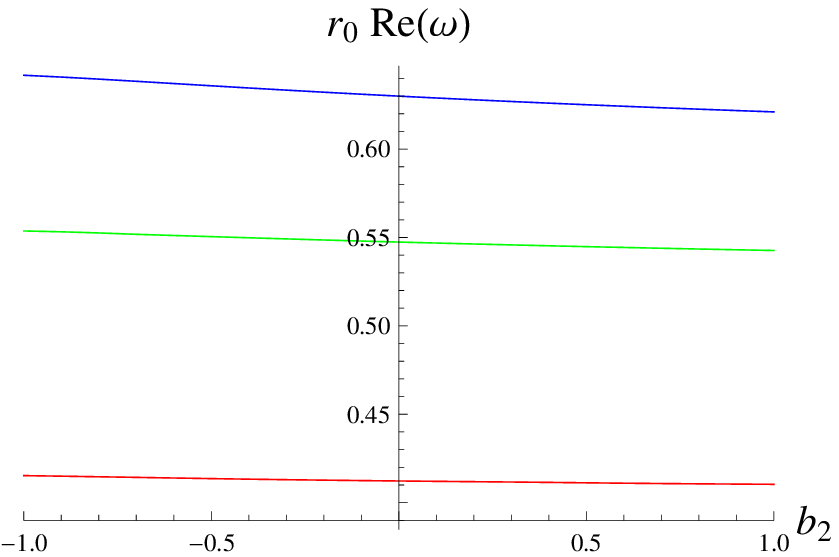}\includegraphics*{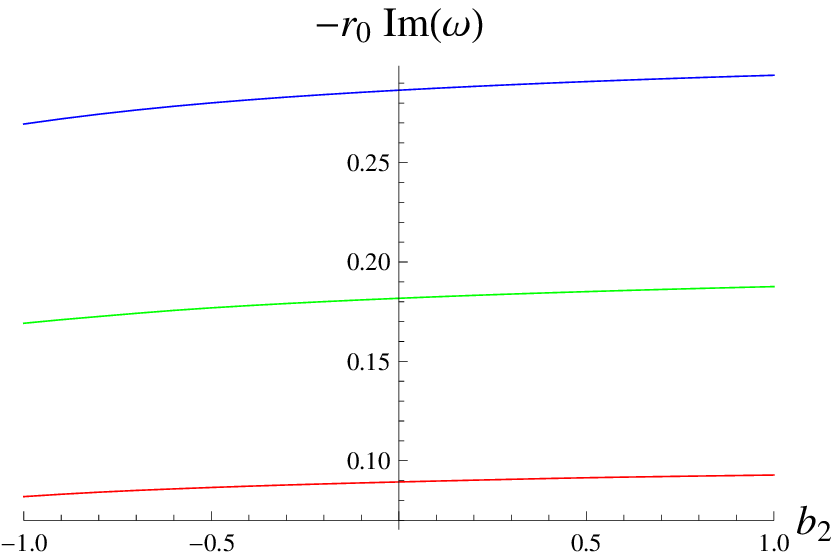}\includegraphics*{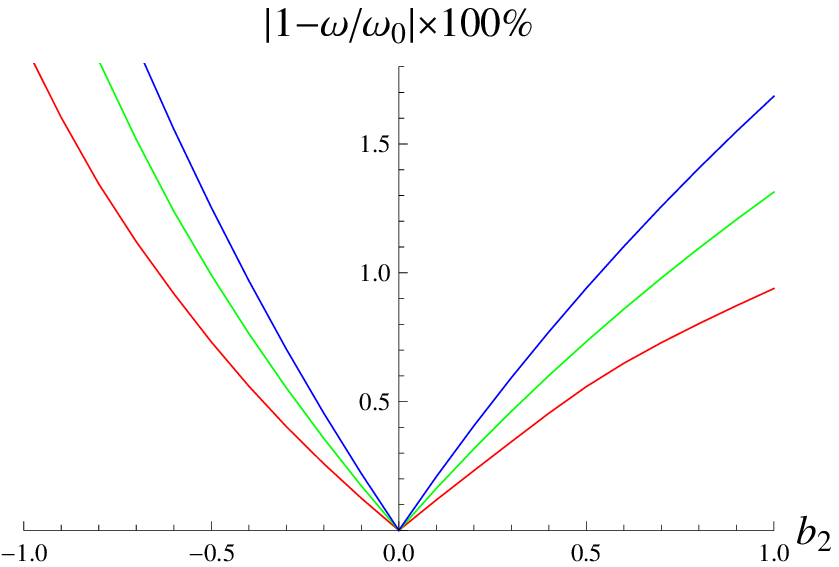}}
\caption{Real and imaginary parts of the fundamental quasinormal mode ($s=1$, $\ell=1$, $n=0$) for $a_{1}=b_{1}=a_{2}=0.5$ and various $\epsilon$ as a function of $b_2$, and the relative deviation from $\omega_0$, corresponding to $b_2=0$: $\epsilon=0.5$ (red), $\epsilon=0$ (green), $\epsilon=-0.5$ (blue).}\label{fig:5}
\end{figure*}

Finally, from figs.~\ref{fig:4}~and~\ref{fig:5} one can immediately see that the coefficients $a_2$ and $b_2$ can correct the value of the quasinormal mode by only a few percents and, if these coefficients are not seemingly larger than the $\epsilon$, $a_1$ and $b_1$ (which is usually the case for a great number of known black-hole solutions \cite{Konoplya:2020hyk}) the correction allowed by coefficients $a_2$ and $b_2$ stay within one percent for the fundamental mode and within a couple of percent for the first overtone. Although the higher overtones are generally more sensitive to the small changes of the metric and, thereby, to the higher coefficients, these modes are less relevant for the gravitational-wave signal analysis.

From the above data we conclude that there is indeed a strict hierarchy of the coefficients of the parametrization, and the higher coefficients are much less important than the lower ones when determining dominant quasinormal modes. If we want to describe low-lying quasinormal frequencies with the relative error of about one-two percents, it is sufficient to include only the three parameters $\epsilon$, $a_1$ and $b_1$, provided that the evident constraints on the compactness are imposed.

\section{Quasinormal ringing and echoes of nonmoderate black holes}\label{sec:nonmoderate}

\begin{figure}
\centerline{\resizebox{\linewidth}{!}{\includegraphics*{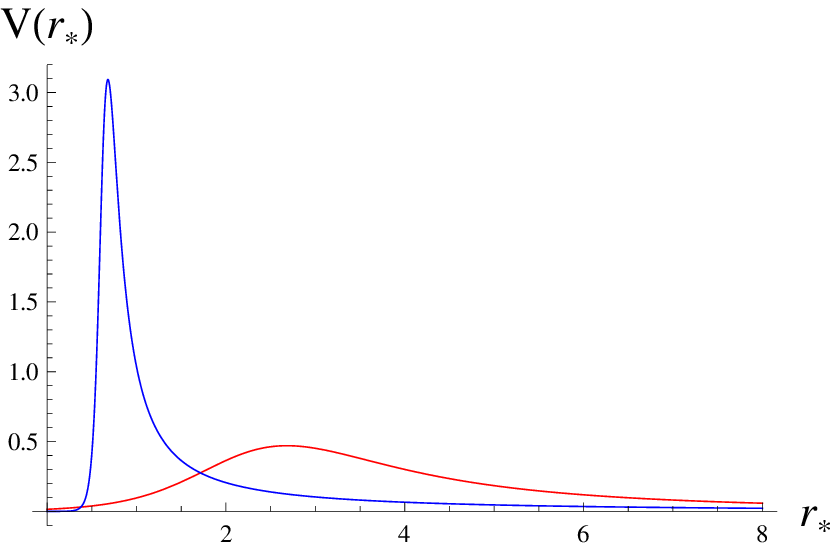}}}
\centerline{\resizebox{\linewidth}{!}{\includegraphics*{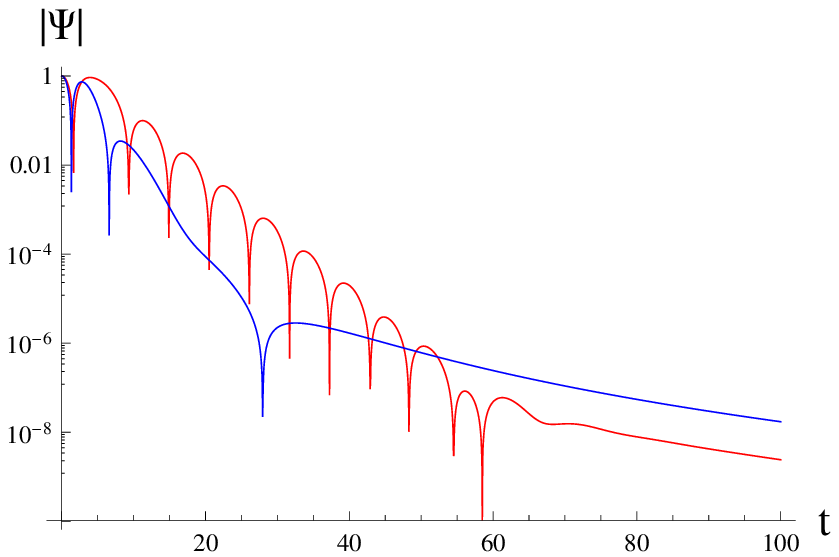}}}
\caption{Effective potentials and time-domain profiles ($s=1$, $\ell=1$, $r_0=1$) for $b_{1}=a_{2}=b_{2}=\ldots=0$ and $\epsilon=-0.5$: $a_{1}=0$ ($\omega\approx0.562-0.301\imo$, red) and $a_{1}=20$ ($\omega\approx0.3-0.6\imo$, blue).}\label{fig:6}
\end{figure}

Here we will consider the other situation, when the higher coefficients are much larger than the lower ones. In Fig.~\ref{fig:6} one can see the effective potential of the electromagnetic field for $\epsilon=-0.5$ in two particular cases: one for which all other coefficients are equal to zero and the other when $a_1=20$. The first case is characterized by a relatively slowly changing effective potential, while the second case is represented by a distinctive, very high and narrow peak, which is quite close to the event horizon. Generally speaking, the strong change of the metric function in the radiation zone could lead either to a high peak of the effective potential or a very deep negative gap near the event horizon. The latter would almost definitely signify the existence of a bound state with negative energy, causing instability of the perturbations. Therefore, the typical picture of the effective potential for a stable perturbation is a distinctive high peak.

Such a distinctive behavior of the effective potential results in an enormously strong deviation of the quasinormal modes from their Schwarzschild values. Thus, on Fig.~\ref{fig:6} we see that the quasinormal modes of the electromagnetic perturbations for a nonmoderate geometry are characterized by much a smaller real oscillation frequency and a much higher damping, and the deviation is of the order of hundreds of percent. Such a strong deviation from the Einstein theory is not only beyond the allowed range in observations of gravitational waves from black holes~\cite{alternative1}, but, in extreme cases, would even show itself as violations of the known observational constraints in the post-Newtonian regime.

\begin{figure}
\centerline{\resizebox{\linewidth}{!}{\includegraphics*{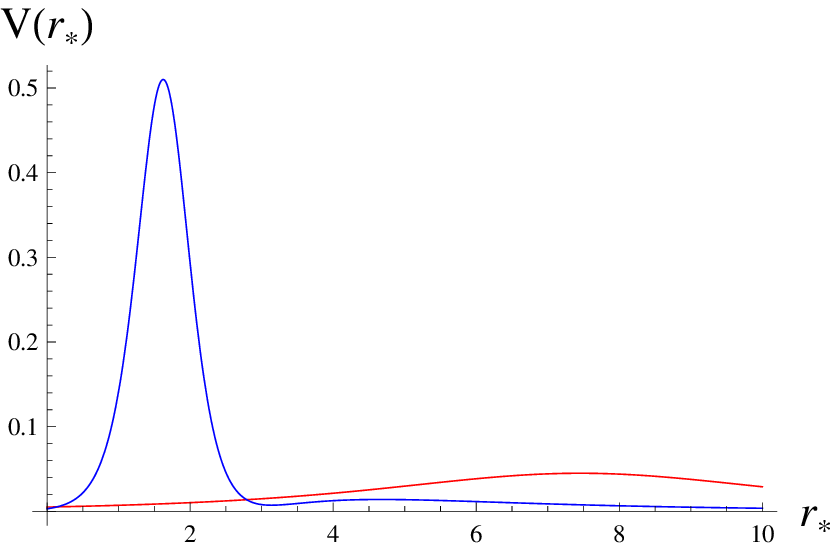}}}
\centerline{\resizebox{\linewidth}{!}{\includegraphics*{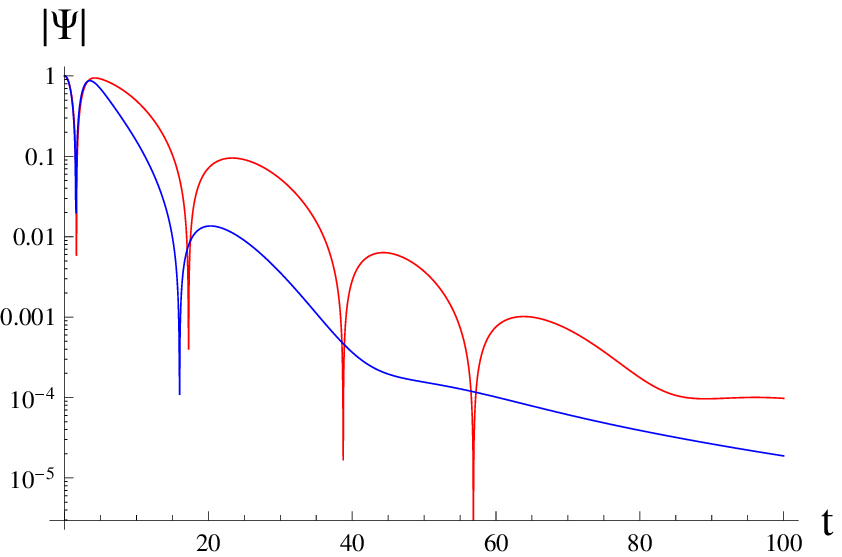}}}
\caption{Effective potentials and time-domain profiles ($s=0$, $\ell=0$, $r_0=1$) for $b_{1}=a_{2}=b_{2}=\ldots=0$ and $\epsilon=0.5$: $a_{1}=0$ ($\omega\approx0.153-0.120\imo$, red) and $a_{1}=4$ ($\omega\approx0.13-0.21\imo$, blue).}\label{fig:7}
\end{figure}

From Fig.~\ref{fig:7} we see that for the fundamental scalar-field mode a large deviation from the Schwarzschild behavior manifests itself for even smaller values of $a_{1}$.

\begin{figure}
\centerline{\resizebox{\linewidth}{!}{\includegraphics*{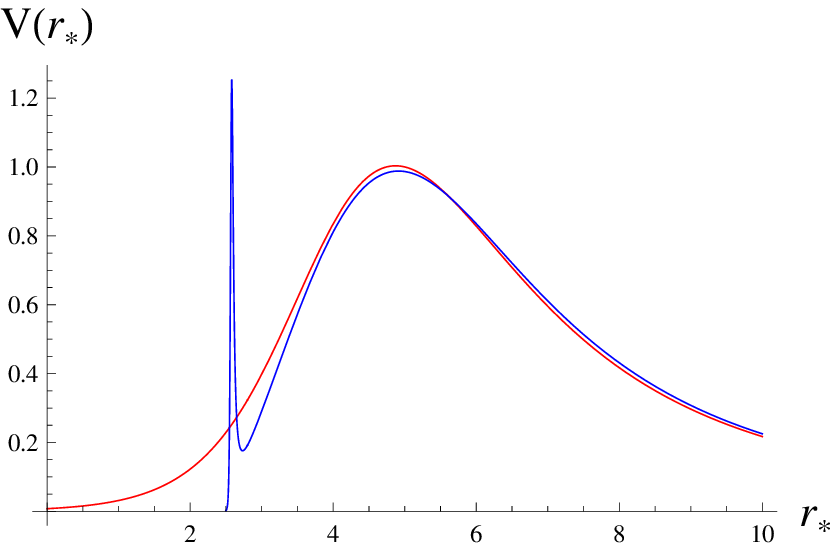}}}
\centerline{\resizebox{\linewidth}{!}{\includegraphics*{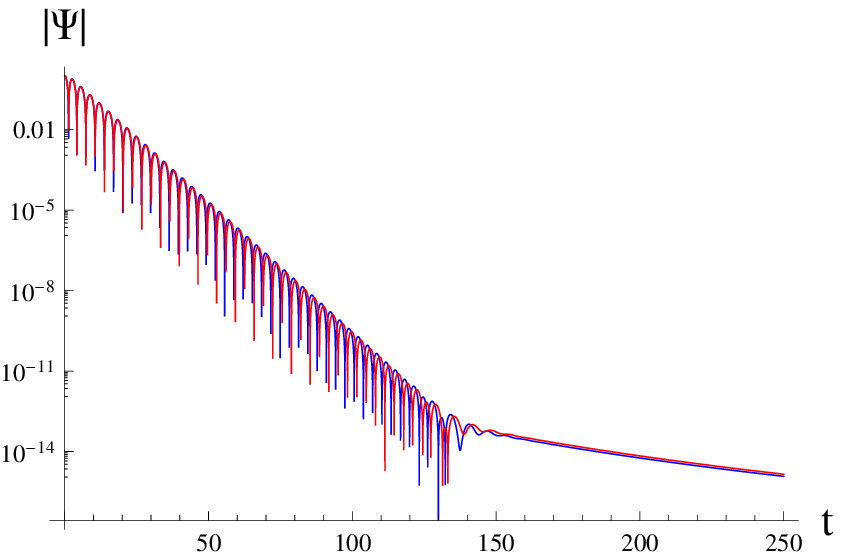}}}
\caption{Effective potentials and time-domain profiles ($s=0$, $\ell=2$, $r_0=1$) for $\epsilon=a_{1}=a_{2}=b_{2}=\ldots=0$, $b_{1}=-0.3$ ($\omega\approx0.9667-0.2237\imo$, red) and $\epsilon=a_{1}=a_{2}=b_{3}=\ldots=0$, $b_{1}=-0.99$ and $b_{2}=10$ ($\omega\approx0.9762-0.2230\imo$, blue).}\label{fig:8}
\end{figure}

\begin{figure}
\centerline{\resizebox{\linewidth}{!}{\includegraphics*{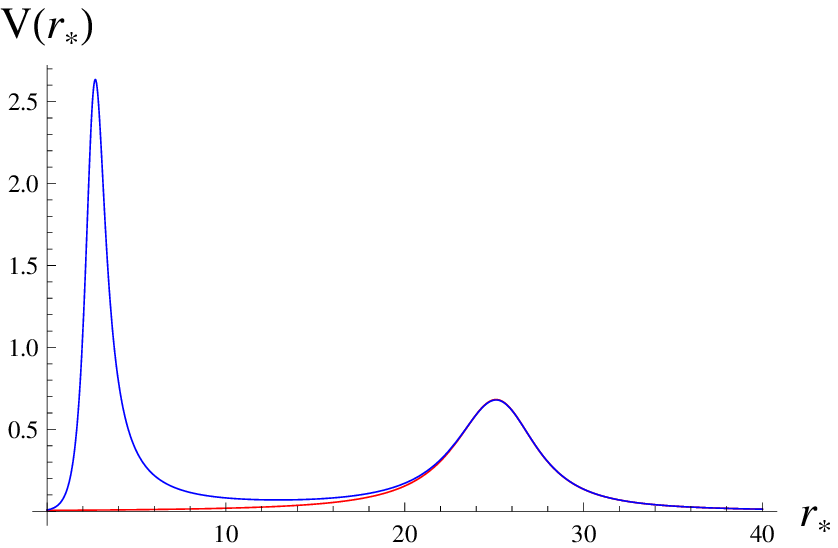}}}
\centerline{\resizebox{\linewidth}{!}{\includegraphics*{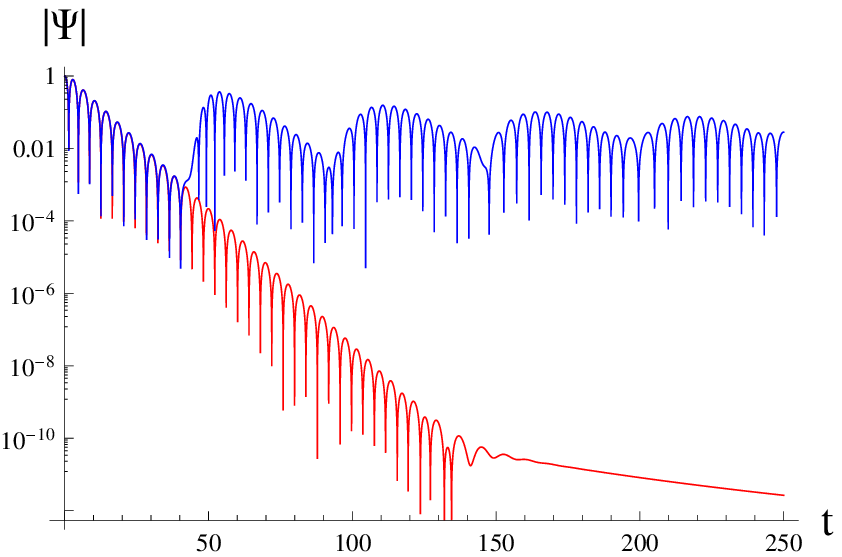}}}
\caption{Effective potentials and time-domain profiles ($s=0$, $\ell=1$, $M=0.5$) for $r_0=0.435$, $\epsilon = 1.3$, $a_{0} = 1.5$, $a_{1} = 0.13$, $a_{2}=0$ $b_{1}=b_{2}=\ldots=0$ ($\omega\approx0.791-0.172\imo$, red) and $r_0 = 0.2$, $\epsilon=4$, $a_{0}=7$, $a_{1} =0.5$, $a_{2} = -8$, $a_{3} = 8$, $a_{4} = -0.1$, $a_{5}=0$, $b_{0}=b_{1}=b_{2}=\ldots=0$ (blue, echo).}\label{fig:9}
\end{figure}

The other case of nonmoderate black-hole spacetimes is the black-hole geometry which looks as Schwarzschildian everywhere, except for a very small region near the event horizon. Such geometries do not change the classical radiation processes, such as quasinormal modes, shadows, accretion of matter, etc., but at very late times produce additional, and still elusive, scattering of gravitational waves called \emph{echoes} \cite{Cardoso:2016rao}.
Then, the nonmoderate geometry can be different from some moderate Schwarzschild-like black hole in such a small region that it does not impact (significantly) the time-domain profile. From Fig.~\ref{fig:8} we see that the additional peak of the effective potential is so narrow that it allows the signal to tunnel into the horizon rather than reflect to infinity causing no observable echo in the time-domain profile and almost no change in the ring-down phase. However, once the additional peak near the effective potential is broader, it can produce echoes, and still will not significantly influence the quasinormal ringing and other astrophysically relevant observables (see Fig.~\ref{fig:9}).

\section{Conclusions}

Here we considered low-lying quasinormal modes for the general parametrized spherically symmetric black hole suggested in \cite{Rezzolla:2014mua}. We have shown that there is a strong hierarchy of the coefficients of the parametrization such that every order of expansion is roughly one order less important. This way, if one wants to determine quasinormal modes with the relative error of about one to two percent, only three coefficients are sufficient provided that the evident constraints on the compactness are imposed and the metric changes moderately in the radiation zone between the event horizon and the innermost stable circular orbit.
The case of nonmoderate stable black-hole spacetime is characterized by a high distinctive peak of the effective potential near the event horizon, which leads either to an enormous deviation in quasinormal frequencies from their Einsteinian values by hundreds of percent or to a Schwarzschild-like metric which changes strongly only in a tiny zone in the vicinity of the event horizon, resulting in a Schwarzschild-like quasinormal modes accompanied by echoes at very late times.

Therefore, we conclude that astrophysically observable quantities can be very well described by the general parametrization of the black-hole spacetime truncated at the first or, at most, second order of the continued-fraction expansion.

Our work can be extended to the case of rotating black holes, at least once the perturbation equations allow for the separation of variables \cite{Konoplya:2018arm}.

\acknowledgments
A.~Z. was supported by Conselho Nacional de Desenvolvimento Científico e Tecnológico (CNPq). R. K. would like to acknowledge support of the Grant N\textsuperscript{\underline{o}}~19-03950S of Czech Science Foundation (GAČR).

\end{document}